# A Giant Planet Imaged in the Disk of the Young Star β Pictoris

A.-M. Lagrange,[1]* M. Bonnefoy,[1] G. Chauvin,[1] D. Apai,[2] D. Ehrenreich,[1] A. Boccaletti,[3] D. Gratadour,[3] D. Rouan,[3] D. Mouillet,[1] S. Lacour,[3] M. Kasper[4]

[1]Laboratoire d'Astrophysique, Observatoire de Grenoble, Université Joseph Fourier, CNRS, BP 53, F-38041 Grenoble, France. [2]Space Telesope Science Institute, 3700 San Martin Drive, Baltimore, MD 21218, USA. [3]LESIA, UMR 8109 CNRS, Observatoire de Paris, UPMC, Université Paris-Diderot, 5 place J. Janssen, 92195 Meudon, France. [4]ESO, Karl Schwarzschild St, 2, D-85748 Garching bei München, Germany.

*To whom correspondence should be addressed. E-mail: anne-marie.lagrange@obs.ujf-grenoble.fr

**Here we show that the ~10 Myr β Pictoris system hosts a massive giant planet, β Pictoris b, located 8 to 15 AU from the star. This result confirms that gas giant planets form rapidly within disks and validates the use of disk structures as fingerprints of embedded planets. Among the few planets already imaged, β Pictoris b is the closest to its parent star. Its short period could allow recording the full orbit within 17 years.**

Gas giant planets form from dusty gas-rich disks that surround young stars through processes that are not completely understood. Two general mechanisms of such planets have been identified (*1*): disk fragmentation and accretion of gas onto a solid, typically 5-10 Earth-mass (MEarth) core. Currently, available models do not offer a detailed description of all the physical and dynamical steps involved in these processes. The lifetime of gas-rich disks limits the availability of nebular gas and, thus, defines the time window in which gas giant planets can form. Once formed, giant planets are predicted to interact with the disk and distort it, possibly leading to characteristic disk structures that can be used to infer the presence of planets and to constrain their orbits. Up to now, most giant planets have been detected around stars orders-of-magnitude older than the lifetime of gas-rich circumstellar disks, preventing the validation of models of disk–planet interactions and the final phases of giant planet accretion.

The young (~$12^{+8}_{-4}$ Myr), nearby (19.3 ± 0.2 pc), 1.75 solar mass (MSun) star β Pictoris (*2*, *3*) hosts a wide (several hundreds of AUs), tenuous edge-on circumstellar dust disk (*4*). It is composed of dust particles continuously replenished through collisions of larger solid bodies (planetesimals, comets), and is referred to as a debris disk (*5*, *6*), in contrast to more massive gas-rich counterparts around younger (a few Myr) stars. This disk has been studied in great detail over the past 25 years. Observations at optical to the thermal infrared wavelengths revealed multiple disk structures (*7–9*), as well as asymmetries in disk size, scale height, and surface brightness distributions (*10*, *11*).

Some of these structures and asymmetries have been theoretically linked to the presence of one or more massive planets. An inner warp in the disk plane (*12*, *13*), in particular, can be reproduced by detailed models that include a planetary-mass companion (*13*, *14*). In addition, spectroscopic observations over several years revealed sporadic high-velocity infall of ionized gas to the star, attributed to the evaporation of comet-like bodies grazing the star [see, *e.g.* (*5*, *15*, *16*)]. The observed comet infall has been attributed to the gravitational perturbations by a giant planet within the disk [see, *e.g.* (*17*), and references therein]. Taken together, data and models suggest that the β Pictoris disk is populated by dust, gas, solid kilometer-sized bodies, and possibly planet(s).

Near-infrared images of β Pictoris obtained in 2003 (*18*) show a faint (apparent magnitude $L' = 11.2$ mag), point-like source at ~8 AU in projected separation from the star, within the North-East side of the dust disk. However, these data were not sufficient to determine whether this source was a gravitationally bound companion, or an unrelated background star, whose projected position in the plane of the sky happened to be close to β Pictoris. Further observations in January and February 2009 did not detect the companion candidate (*19*, *20*), an outcome fully consistent with both the proper motion of β Pictoris with respect to a background star or with the orbital motion of a physically bound companion.

Here we present high-contrast and high-spatial resolution near-infrared images obtained in October, November, and December 2009 with the European Southern Observatory's Very Large Telescope's (VLT) Adaptive Optics NaCo instrument (*21*, *22*) (see SOM for more details on the observations and data reduction). The images obtained in October 2009 (Fig. 1) show a faint point source South-West of the star, with a brightness ($\Delta L' = L^* - L = 7.8 \pm 0.3$) comparable to that ($\Delta L' = L^* - L = 7.7 \pm 0.3$) of the source detected North-East of β Pictoris in November 2003 (Fig. 1).

The source lies at a projected separation of 297.6 ± 16.3 mas, and at a position angle (PA) of 210.6 ± 3.6 degrees. Within the error bars, the source is located in the plane of the disk. To confirm the signal detected SW of β Pictoris in October 2009, we gathered further data in November and December 2009. Together, these data confirm the detection made in October 2009 (see SOM).

The images show that the source detected in November 2003 could not have been a background object (Fig. 2). Indeed, if background, given the star's proper motion (table S1, SOM), the Nov. 2003 source would be located and detectable 5.1 AU away, South-East (PA = 147.5 deg) of β Pictoris in Fall 2009. The data do not show such a source (fig. S2). On the contrary, the source position in Fall 2009 is compatible with the projected position in November 2003 if the source is gravitationally bound to the star (see below).

Based on the system age, distance, and on the apparent brightness of the companion, the widely-used Baraffe *et al.* (*23*) evolutionary models predict a mass of ~9 ± 3 Jovian masses (MJup). This value is compatible with those derived from other groups [see, *e.g.* (*24*)] calculations when assuming—as Baraffe *et al.* (*23*)—that the planets form from the spherical contraction and cooling of a hot, initially non-rotating cloud of gas. However, it is not clear whether the basic assumption of these "hot start" models, *i.e.*, the spherical contraction, applies to β Pictoris b or other planets. A leading alternative model (*24*), recently developed, includes the loss of energy of the infalling gas via an accretion shock (core accretion start), resulting in the accretion of cooler gas to the forming planet. This "cold start" model predicts luminosities two orders of magnitudes lower than those predicted by the "hot start" models at young (~10 Myr) ages for a ~10 MJup mass planet. The initial difference between the two models decreases after formation; it remains however significant for these massive planets even at ages of 100 Myr (factor of 10). The low luminosities predicted by the" cold start" model are not easy to reconcile with the brightness of β Pictoris b. This apparent inconsistency suggests that this model overestimates the energy lost during the formation of beta Pictoris b. The assumptions underlying the "hot start" or "cold start" models are still a matter of debate, and observations of objects such as β Pictoris b are essential to test them. In what follows, we use the mass inferred by the "hot start" model because there are lines of evidence that the companion cannot be much more massive (see below).

In order to constrain the orbital parameters of β Pictoris b, we took the projected separation measured in November 2003 and computed the expected position of the planet in 2009 (SOM), assuming that it moves in a prograde [following (*25*)], circular orbit within the disk or close to the plane of the disk. Comparison between the expected projected separation in December 2009 and the observed position implies that the planet's semi-major axis is between 8 and 13 AU (fig. S3). This leads to orbital periods of 17 to 35 yr for the planet. The semi-major axis is less strongly constrained in the case of an eccentric orbit because of the unknown longitude of the periastron. However, for the probable case of a moderate eccentricity [$e < 0.05$ (*26*)], the semi-major axis must be in the range 8 to 15 AU. These orbital parameters are compatible with the non-detection at $L'$-band in February 2009 (Fig. 2), given the 4-sigma detection limits at this date which correspond to projected separations of 6.5 AU. Thus, β Pictoris b orbits closer to its parent star than Uranus and Neptune do in the Solar System.

The planet separation is qualitatively consistent with the observation of belt-like structures in the inner disk at 6 ± 3 and 16 ± 3 AU (*7*, *8*). The separation and mass are fully consistent with those predicted by dynamical studies that invoked a planet to reproduce the inner disk warp (*13*, *14*). A more massive (>40 MJup) companion at such separations would on the contrary, not be compatible with the warp constraints (SOM).

Finally, it has been suggested (*27*) that the 2003 candidate planet could have been responsible for a peculiar photometric variability event observed in 1981, when transiting in front of the star. The 2009 data are compatible with this possibility [see also (*19*)], however, only for a small range of planet orbital parameters.

The detection of β Pictoris b follows on the recent detections of planets around the intermediate-mass stars HR 8799 and Fomalhaut [see SOM and (*28*, *29*)]. These stars are also surrounded by debris disks (*30*, *31*). However, β Pictoris b orbits closer to its star and is younger than the planets around HR 8799 and Fomalhaut (30 to 160 and 100 to 300 Myr old, respectively, tables S2A and S2B). Our images of β Pictoris b provide direct evidence that massive giant planets can form rapidly, on time scales of a few million years within circumstellar disks (*32*). This is in agreement with studies of the dispersion of primordial disks around young intermediate mass stars, which yield typical disk lifetimes of between less than 3 and 6 Myr (*16*).

A comparison of the luminosity of β Pictoris [8.7 LSun (*3*)] to that of the Sun suggests that the orbit of β Pictoris b lies at or slightly beyond the disk radius outside which water is stable as ice (snow-line). The snowline is thought to separate disk regions where rocky or gaseous/icy planets form (*33*, *34*). Indeed, beyond the snowline, the disk surface density is expected to be higher (factor of 3) than that inside the line; this allows giant planet cores (10 MEarth) to form before the dispersion of the gaseous nebulae. Core accretion models suggest indeed that this latter step—the onset of rapid gas accretion before the loss of circumstellar gas—is the critical step in forming giant planets. Assuming a core with a minimum mass of 5-10 MEarth and taking into account the snowline properties as a function of stellar mass and age, the model of Kennedy and Kenyon (*33*) determines the snow line position as a function of time, and possible locations of giant protoplanet cores as a function of stellar mass. For a 2 MSun star, the snowline location varies between 2.5 and 4 AU for ages between 1 and 10 Myr, respectively. In the case of the 1.75 MSun β Pictoris, core-accretion based models predict a rapid formation of giant protoplanet cores between ~6 and 18 AU. The observed orbital radius of β Pictoris b is consistent with this range, demonstrating that the planet could have formed via core accretion on the same orbit where it is observed today. This possibility is in contrast to the case of the more distant planets Fomalhaut, HR8799bc, AB Pic b, and 2MASS 1207b, which are too massive (tables S2A and S2B) to have formed at their present separations (40 AU or larger), via core accretion.

35. We thank ESO staff, especially C. Melo, C. Dumas, and J. Girard for their help. We acknowledge financial support from the Programme National de Planétologie (INSU), as well as from the Agence Nationale pour la Recherche (grant NT05-4_4463). D. E. acknowledges support from the Centre National d'Etudes Spatiales. We also thank H. Beust, P. Rubini, and D. Collardey.

**Supporting Online Material**
www.sciencemag.org/cgi/content/full/science.1187187/DC1
SOM Text
Figs. S1 to S4
Tables S1 and S2
References and Notes

**Fig. 1.** β Pictoris imaged at *L'*-band (3.78 microns) with the VLT/NaCo instrument in November 2003 (**left**) and in the fall of 2009 (**right**). We used images of the comparison star HR2435 to estimate and remove the stellar halo (SOM). Similar results are obtained when using Angular Differential Imaging (SOM).

**Fig. 2.** Expected positions (spiraling red curve), between November 2003 (green circle) and December 2009, of the source detected in November 2003 if it was a background object with a projected position at this epoch close to β Pictoris. We computed these positions using the proper motion of β Pictoris. Epochs of interest are indicated by open red circles. The measured positions of the detected sources in November 2003 and in Fall 2009 are indicated by respectively the green and blue filled circles). The gray filled circle indicates the expected position of the planet in February 2009, assuming a semi major axis of 8 AU and a circular orbit. The planet position at less than 4 AU from the star, was well below the detection limit of the February 2009 data.

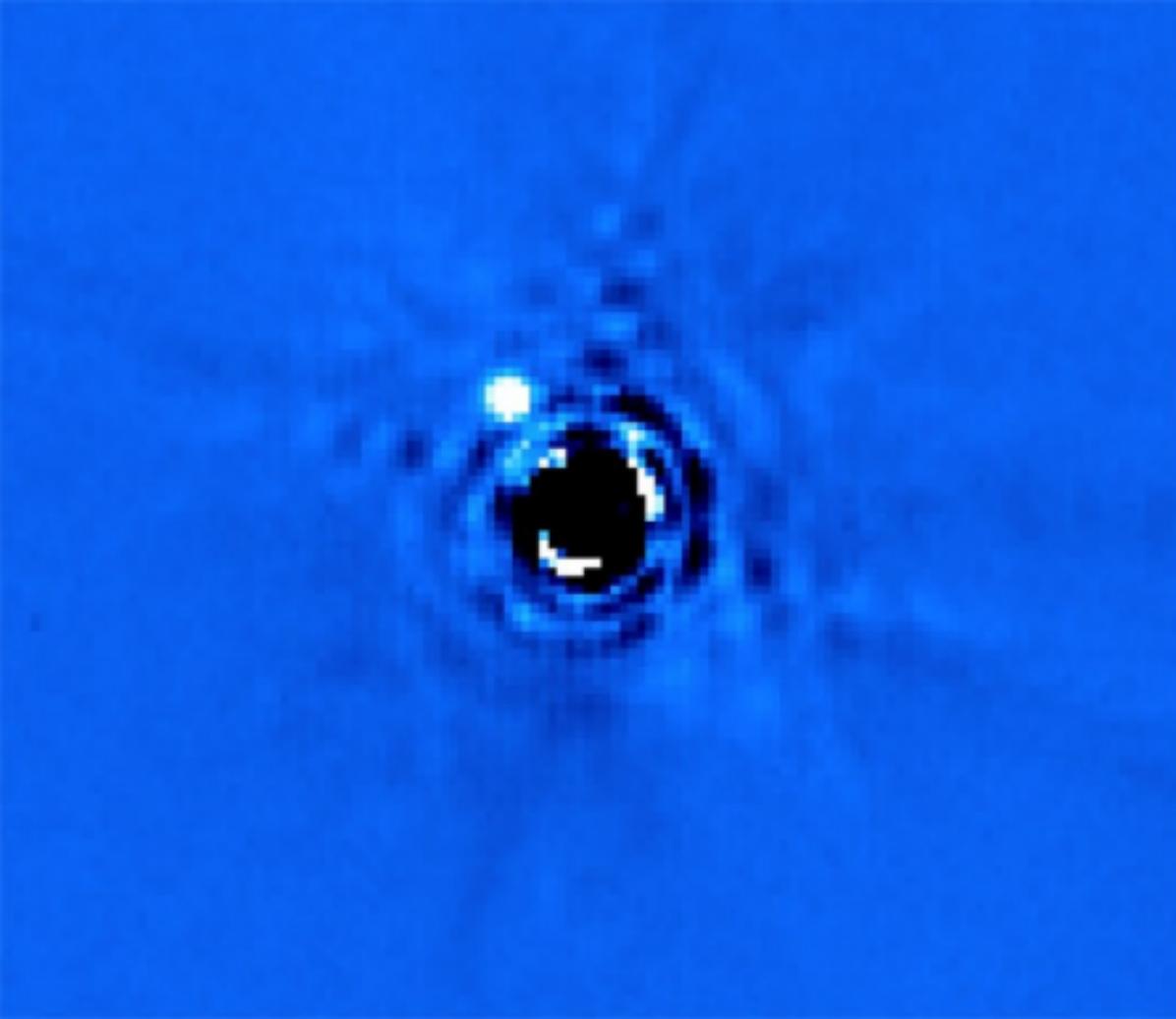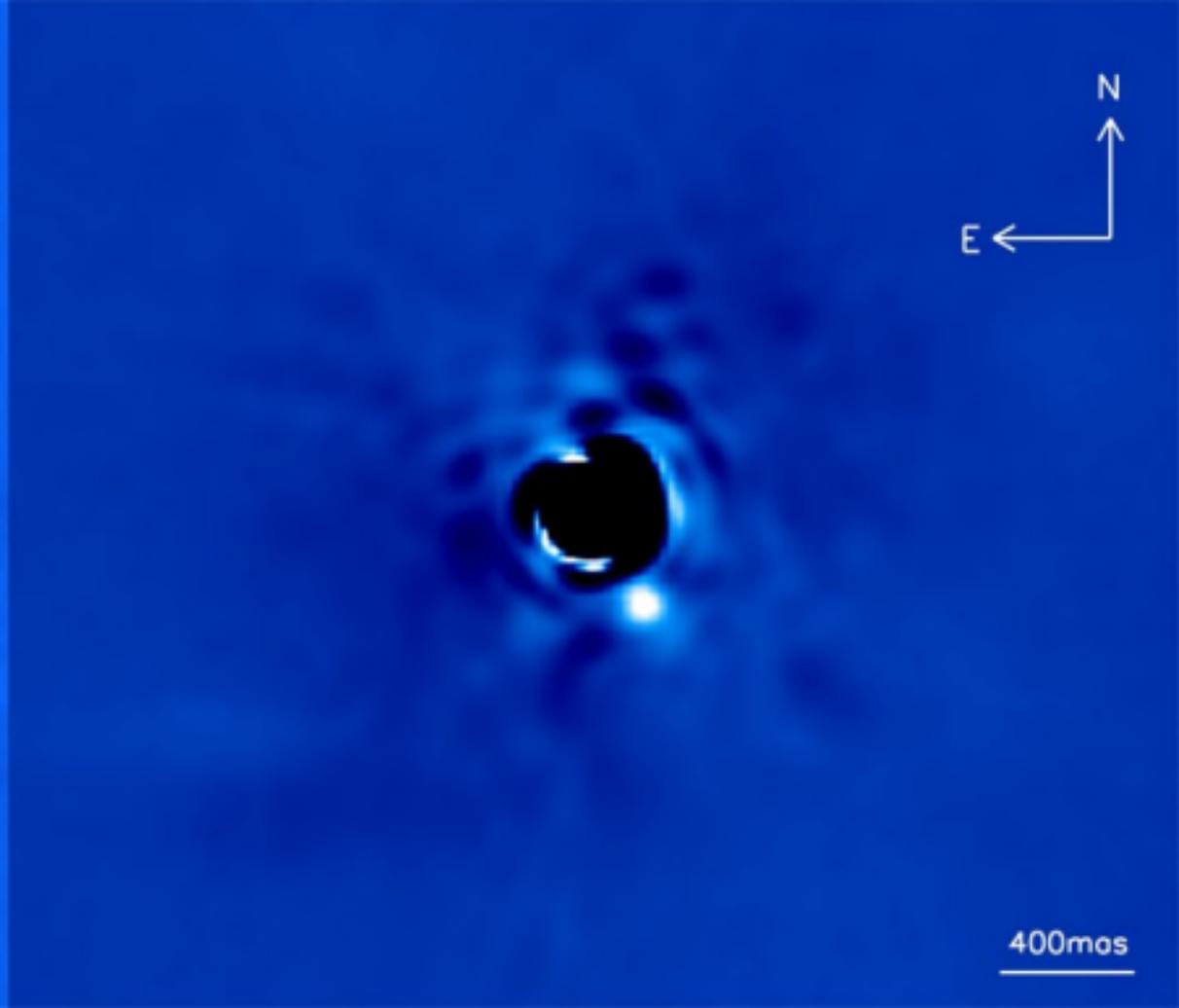

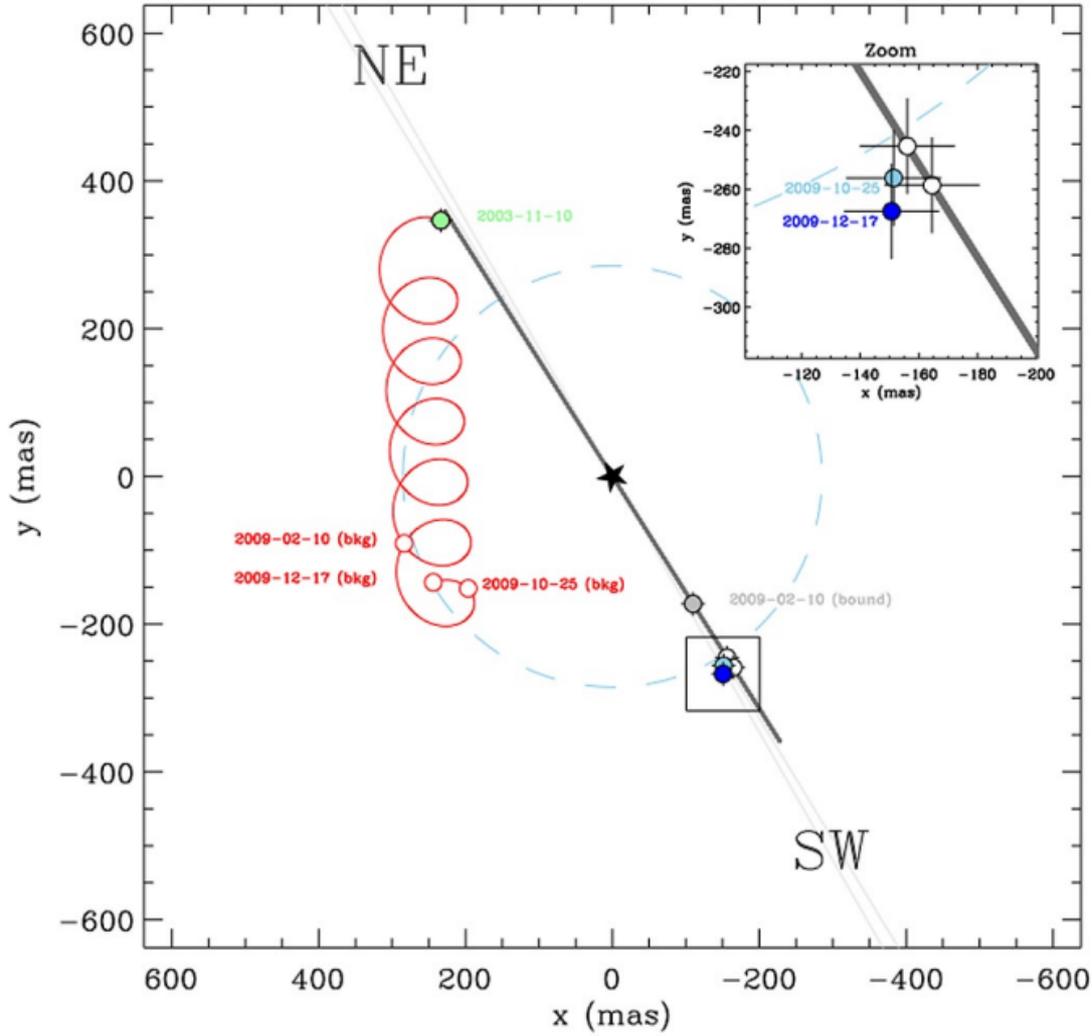